\documentclass[twocolumn,showpacs,amsmath,amssymb,floatfix,prb]{revtex4}

\usepackage{graphicx}
\begin{document}

\title{Contractor-Renormalization approach to frustrated magnets in a magnetic field}
\author{A. Abendschein}
\email{abendschein@irsamc.ups-tlse.fr} 
\affiliation{Laboratoire de Physique Th\'eorique, Universit\'e Paul Sabatier, CNRS, 31062  Toulouse, France}
\author{S. Capponi}
\affiliation{Laboratoire de Physique Th\'eorique, Universit\'e Paul Sabatier, CNRS, 31062  Toulouse, France}

\begin{abstract}
We propose to use the Contractor Renormalization (CORE) technique in
order to derive effective models for quantum magnets in a magnetic
field. CORE is a powerful non-perturbative technique that can reduce
the complexity of a given microscopic model by focusing on the
low-energy part.  We provide a detailed analysis of frustrated spin
ladders which have been widely studied in the past: in particular, we
discuss how to choose the building block and emphasize the use of
their reduced density matrix. With a good choice of basis, CORE is
able to reproduce the existence or not of magnetization plateaux in
the whole phase diagram contrary to usual perturbation theory. 
We also address the issue of plateau formation in two-dimensional bilayers and point out the analogy between
non-frustrated strongly anisotropic models and frustrated SU(2) ones. 

\end{abstract}

\date{\today}
\pacs{75.10.Jm, 75.60.Ej}

\maketitle

\section{Introduction}
\label{introduction.sec}

In the presence of a magnetic field, quantum magnets exhibit
fascinating properties. In particular, it can happen that the uniform
magnetization along the field exhibits plateaux for rational values,
which has given rise to lots of
theoretical~\cite{Oshikawa1997,Totsuka1998a,Cabra1998,Honecker2004}
and experimental~\cite{Kageyama1999,Kikuchi2005b,Narumi2004a} work. 
The appearance of such plateaux was found to be favored by magnetic
frustration.

More recently, experiments on spin dimer compounds (such as
TlCuCl$_3$, KCuCl$_3$ and
BaCuSi$_2$O$_6$, see Ref.~\onlinecite{Cavadini2002b,Oosawa2002a,Ruegg2003,Jaime2004})
have shown that the triplet excitations behave as bosons that can form
superfluid (absence of plateau) or crystalline (finite plateau) phases
depending on the competition between repulsive and kinetic
interactions. Moreover, the possibility of having both orders, namely
a supersolid, could potentially be observed in related compounds. As
is well-known, frustration reduces triplet delocalization and thus, is
favorable to solid behaviour, i.e. plateau formation, or supersolid behaviour.
However, frustrated spin models are difficult to study numerically due
to the sign-problem of the Quantum Monte-Carlo (QMC) technique and the
absence of reliable large-scale numerical techniques in two dimensions
or higher. However, the effective bosonic models themselves can often
be simulated when the frustration has disappeared and is absorbed in
the effective parameters; indeed, such effective bosonic models have
been proposed either based on perturbation
theory~\cite{Totsuka1998a,Mila1998,Momoi2000} or on phenomenological
grounds.~\cite{Bendjama2005} Therefore, we think that it would be
highly desirable to derive non-perturbative effective parameters
directly from microscopic models and we propose to use the
contractor-renormalization (CORE) technique to do so.

The CORE method has been proposed in
Ref.~\onlinecite{Morningstar1994,Morningstar1996,Weinstein2001} as a
systematic algorithm to derive effective Hamiltonians and operators
that contain all the low-energy physics. In principle, these effective
operators are given by an infinite cluster expansion. CORE has been
successfully applied to a variety of both
magnetic~\cite{Piekarewicz1997,Weinstein2001,Berg2003,Capponi2004,Budnik2004,Li2005}
and doped~\cite{Altman2002,Capponi2002} low-dimensional systems and it
turns out that, in most cases, the cluster expansion converges quite
fast, which is a necessary condition for any practical implementation
of this algorithm. Still, the issue about CORE convergence is crucial
and currently debated.~\cite{Siu2006}

In section~\ref{ladder.sec}, we remind the reader of the CORE
algorithm and investigate the frustrated 2-leg ladder. Being a
well-known model, we can compare our findings to other
well-established numerical results and discuss the accuracy of CORE as
well as its convergence. In section~\ref{choicebasis.sec}, we
investigate how to choose the best block decomposition and how to
select the low-lying states to keep by using information obtained
with the exact reduced density-matrix of a block embedded in a large
system. Finally, in section~\ref{2d.sec}, we turn to some
two-dimensional (2D) bilayer spin models which are candidates for
observing some of these exotic bosonic phases.

\section{Frustrated 2-leg ladder}
\label{ladder.sec}

\begin{figure}
\begin{center}
\includegraphics[width=.48\textwidth,clip]{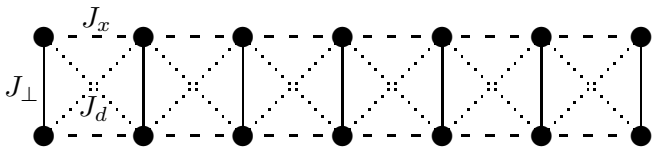}
\end{center}
\caption[]{The spin ladder with coupling $J_{\perp}$ on vertical rungs, $J_x$ along the legs and $J_d$ on the diagonal bonds.
\label{ladder.fig} }
\end{figure}

The Hamiltonian ${\cal H}$ of the spin $S = \frac {1} {2}$ frustrated
antiferromagnetic Heisenberg ladder in an external magnetic field $h$
reads:
\begin{eqnarray}
\label{heisenberg_hamilton}
  {\cal H} &=& J_{\perp} \sum_{r=1}^{L} {\bf S}_{r,1} \cdot {\bf S}_{r,2} + J_{x} \sum_{r=1}^{L} \sum_{i=1}^{2} {\bf S}_{r,i} \cdot {\bf S}_{r+1,i} + \\
  &+& J_{d} \sum_{r=1}^{L} \big{(} {\bf S}_{r,1} \cdot {\bf S}_{r+1,2} + {\bf S}_{r,2} \cdot {\bf S}_{r+1,1} \big{)} - h \sum_{r=1}^{L} \sum_{i=1}^{2} S_{r,i}^{z}. \nonumber
\end{eqnarray}
In accordance with Fig.~\ref{ladder.fig}, the index $r={1,\ldots,L}$ represents the $L$ different rungs of
the ladder whereas $i={1,2}$ indicates the two chains which constitute
the ladder and we use periodic boundary conditions along the legs. The
rung spin exchange $J_{\perp}$ is set to $1$, while $J_{x}$ and
$J_{d}$ stand for the interactions along the chain and diagonal
respectively, which makes the ladder a frustrated system. Note that
due to symmetry $J_x$ and $J_d$ can be interchanged. The role of frustration in the plateau formation can also be understood in 
a related ladder model.~\cite{Honecker2000}

Throughout this paper, 
we only consider the physical properties at zero temperature and in the presence of a finite magnetic field, 
or more specifically, the 
possibility or not of a magnetization plateau at half-saturation~: $m_z=\frac{1}{2} m_{sat}$.

\subsection{Summary}

\begin{figure}
\begin{center}
\includegraphics[width=.48\textwidth,clip]{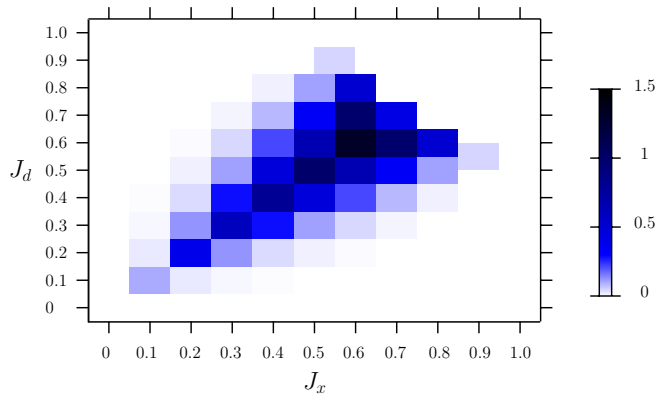}
\end{center}
\caption[]{(Color online) Size of the magnetization plateau at $m_z=\frac {1} {2} m_{sat}$. 
Results have been obtained after finite size scaling analysis 
of exact diagonalization data obtained on $2 \times L$ ladder (up to $L=20$).
\label{size_plat_ED.fig} }
\end{figure}

\begin{figure}
\begin{center}
\includegraphics[width=.48\textwidth,clip]{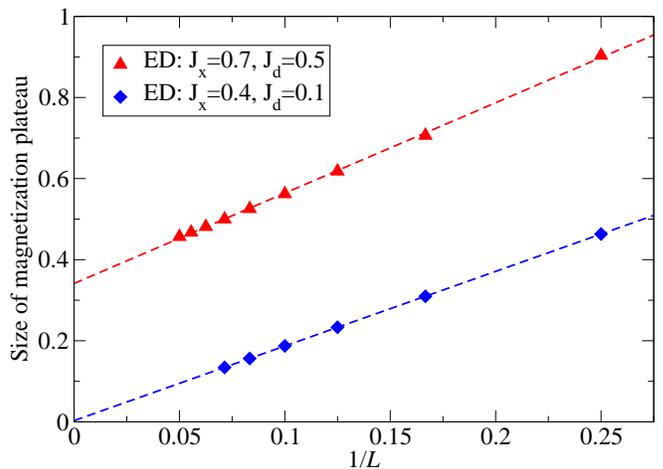}
\end{center}
\caption[]{(Color online) Finite-size scaling of the size of the magnetization plateau at $m_z= \frac {1} {2} m_{sat}$. Data from numerical exact diagonalization is shown for an example of a 
finite plateau ($J_x=0.7$, $J_d=0.5$) and a vanishing plateau ($J_x=0.4$, $J_d=0.1$).}
\label{example_fss.fig}
\end{figure}

The frustrated Heisenberg ladder has been studied with various analytical and numerical 
techniques~\cite{Zheng1998,Allen2000,Wang2000a}. In the absence of magnetic field, there are
two main regions called rung singlet phase and Haldane phase. 
The effect of a magnetic field has been studied with an exact diagonalization (ED) 
technique~\cite{Okazaki2000b} in order to clarify the presence or not of finite-magnetization plateaux. For all the Hamiltonians that we will consider
(including the effective ones), $S^{tot}_z$ is a good quantum number; therefore, it is sufficient to compute the ground-state
energy in all $S_z$ sectors (in the absence of any magnetic field) and
then perform a Legendre transform to get the full magnetization vs field curve $m_z(h)$. 
We have applied the same ED technique and we provide on
Fig.~\ref{size_plat_ED.fig} a sketch of its phase diagram. Data have been extrapolated to the thermodynamic limit after 
standard finite-size scaling analysis of the plateau size (see examples on Fig.~\ref{example_fss.fig}). A large
magnetization plateau phase is found around the strongly frustrated region $J_x \sim
J_d$. We note the existence of different phases without plateau; in particular at large $J_d \sim J_x$, 
there is a first order transition to the so-called Haldane phase.

\begin{figure}
\begin{center}
\includegraphics[width=.48\textwidth,clip]{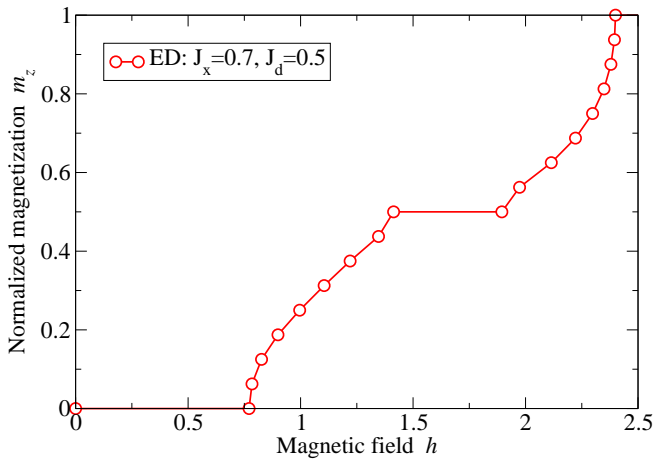}
\end{center}
\caption[]{(Color online) Magnetization $m_z$ along the field (normalized to its saturation value $m_{sat}$) as a function of magnetic field $h$ on a $2 \times 16$ ladder with $J_x=0.7$, and $J_d=0.5$; data from numerical exact diagonalization has been used.}
\label{example_smooth_plat.fig}
\end{figure}

For the plateau phase, we draw on Fig.~\ref{example_smooth_plat.fig} a typical magnetization curve obtained on a $2\times 16$ ladder. 
It exhibits singularities at critical fields~\cite{Affleck1991} and a large half-saturation plateau. Note that 
in order to mimic the thermodynamic limit, we have drawn a line connecting the
middles of the finite-size plateaus of the finite system. 

\subsection{Perturbation theory}
\label{pert_theo.sec}

\begin{figure}
\begin{center}
\includegraphics[width=.48\textwidth,clip]{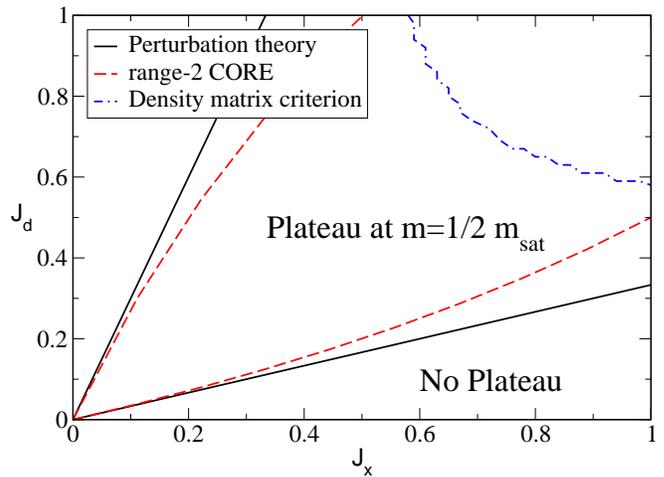}
\end{center}
\caption[]{(Color online) Phase diagram using either perturbation theory or range-2 CORE. Reduced density matrix calculations on a $2 \times 12$ ladder indicate that, for large $J_x$ and $J_d$, the reduced density matrix weight of the 
singlet becomes very small while the $S_z=0$ triplet weight increases substantially 
(region above the blue dashdotted line, see text for details). 
For comparison with the exact results, see Fig.~\ref{size_plat_ED.fig}.}
\label{phase_diag.fig}
\end{figure}

When the only nonzero coupling is $J_\perp$, the ground-state of the
ladder is simply the product of singlets on each rung. The states of a
given rung are labelled as a singlet $|s \rangle_r = \frac {1}
{\sqrt{2}} \big{(} |\uparrow \downarrow \rangle_r - |\downarrow
\uparrow \rangle_r \big{)}$ and the three components of a triplet~:
$|t_{-1} \rangle_r = |\downarrow \downarrow \rangle_r$, $|t_0
\rangle_r = \frac {1} {\sqrt{2}} \big{(} |\uparrow \downarrow
\rangle_r + |\downarrow \uparrow \rangle_r \big{)}$ and $|t_{+1}
\rangle_r = |\uparrow \uparrow \rangle_r$. In the presence of a
finite magnetic field, due to Zeeman splitting, one can restrict the
Hilbert space on each rung to $|s\rangle$ and $|t_{+1}\rangle$, and
then do a perturbation theory~\cite{Mila1998}.  Using pseudo-spin
$S=\frac {1} {2}$ operators $\sigma_r$ for these two states,
Eq. (\ref{heisenberg_hamilton}) can be rewritten as an effective
Hamiltonian $H_{eff}$. Proceeding further, a Jordan-Wigner
transformation leads to a system of interacting, one-dimensional (1D), spinless fermions:
\begin{eqnarray}
\label{ham_sf}
H_{tV} &=& t \sum_{r}^{L} \big{(} c_{r}^{\dagger} c_{r+1} + h. c.) + V \sum_{r}^{L} n_{r} n_{r+1} \nonumber \\
&-& \mu \sum_{r}^{L}n_{r}, 
\end{eqnarray}
where $t$ describes the hopping, $V$ the nearest neighbor interaction
and $\mu$ the chemical potential which can all be expressed in terms
of the previously introduced interactions $J_{\perp}$, $J_x$, and
$J_d$~: 
\begin{eqnarray}
\label{t_V_pert}
t = \frac {J_x - J_d} {2} &\qquad& V = \frac {J_x + J_d} {2} \\
\mu &=& J_{\perp} - h \nonumber
\end{eqnarray}
In the particle language, the occurence or not  of a plateau translates into the existence of single-particle gap  
and the magnetic field plays the role of an effective chemical potential $\mu$ (see Ref.~\onlinecite{Totsuka1998a}). Since the 
metal-insulator
transition of the $t-V$ model~\cite{Haldane1980} occurs at half-filling when $V/|t| = 2$, we conclude that there will be
a finite plateau at half  its saturation value when 
\begin{equation}
\label{pert_theo_inequality}
\frac {1} {3} < J_d / J_x < 3.
\end{equation}
This property is shown in Fig.~\ref{phase_diag.fig} and reasonably agrees with the exact results of Fig.~\ref{size_plat_ED.fig}, 
although we note quantitative differences in the non-perturbative
regime. Therefore, we now turn to a non-pertubative CORE approach with the same basis. 

\subsection{CORE approach in the rung basis}
\label{CORE_app.sec}
The Contractor Renormalization (CORE) method has been formulated in
1994 by Morningstar and
Weinstein~\cite{Morningstar1994,Morningstar1996} and has been used
subsequently to study both magnetic
systems~\cite{Capponi2004,Berg2003,Budnik2004,Weinstein2001,Li2005,Piekarewicz1997}
and doped ones~\cite{Altman2002,Capponi2002,Indergand2006}. The idea
of this non-perturbative method is to derive an effective Hamiltonian
within a truncated basis set which allows to reproduce the low energy
spectrum. This means that the original model is replaced by a model
with fewer states but a more complicated Hamiltonian under the
condition that the retained states of the modified model have an overlap
with the set of lowest lying eigenstates of the full original theory.

For clarity, we briefly remind the main CORE steps and refer to the literature for more details. 
First, one needs to choose a basic cluster and diagonalize it. Then $M$
low-energy states are kept and the remaining states are discarded. 
Generally, the $M$ lowest states are retained, but this is not a necessity as we will discuss in Sec.~\ref{choicebasis.sec}. 
The second CORE step is to diagonalize the full
Hamiltonian ${\cal H}$ on a connected graph consisting of $N_c$ clusters and
obtain its low-energy states $|n \rangle$ with energies $\varepsilon_n$. 
Thirdly, the eigenstates $|n \rangle$ are projected on the tensor
product space of the retained states and Gram-Schmidt orthonormalized
in order to get a basis $|\Psi_n\rangle$ of dimension $M^{N_C}$. Fourthly, the
effective Hamiltonian for this graph is defined as
\begin{equation}
\label{CORE_cluster_def}
h_{N_c} = \sum_{n=1}^{M^{N_c}} \varepsilon_n |\Psi_n \rangle \big{<} \Psi_n|.
\end{equation}
Fifthly, the connected range-$N_c$ interactions $h_{N_c}^{conn}$ can be calculated by substracting the contributions of all connected subclusters. And finally, the effective Hamiltonian is given as a cluster expansion as
\begin{equation}
\label{CORE_ham_def}
H^{CORE} = \sum_{i} h_i + \sum_{<ij>} h_{ij}^{conn} + \sum_{<ijk>} h_{ijk}^{conn} + \cdots .
\end{equation}

Then, of course, one has to study this new effective model, which can
still be a difficult task. One possibility is to iterate CORE
in the renormalization group spirit and study the properties of its
fixed point. Here, following Ref.~[\onlinecite{Capponi2004}], we propose
to check the validity of the effective Hamiltonian \emph{after one
step} by comparing its properties to the exact ones on a given finite
cluster. Of course, once the effective Hamiltonian has been shown to
be accurate, it can be simulated exactly on much larger lattices than
the original model, or thanks to other numerical techniques. For
instance, although the original model is frustrated and cannot be
simulated efficiently by Quantum Monte-Carlo (QMC), there are cases
where the effective Hamiltonian can.

We begin our CORE considerations for the spin $S = \frac {1} {2}$
antiferromagnetic Heisenberg ladder with a rung as simplest possible
basic cluster. 
Because of the Zeeman splitting, and as in the perturbation approach,
we keep only two states per rung~: the singlet and the polarized
triplet $|t_{+1}\rangle$. Then, we have computed up to range-8
effective CORE interactions by solving exactly up to 8 rungs. Despite
having an infinite cluster expansion in Eq.~(\ref{CORE_ham_def}),
previous studies have shown that, in many cases, the long-range
effective interactions decay quickly so that they can be neglected
beyond a certain range $r$.  This is a necessary condition for any
practical implementation of CORE and should be checked systematically.

\begin{figure}
\begin{center}
\includegraphics[width=.48\textwidth,clip]{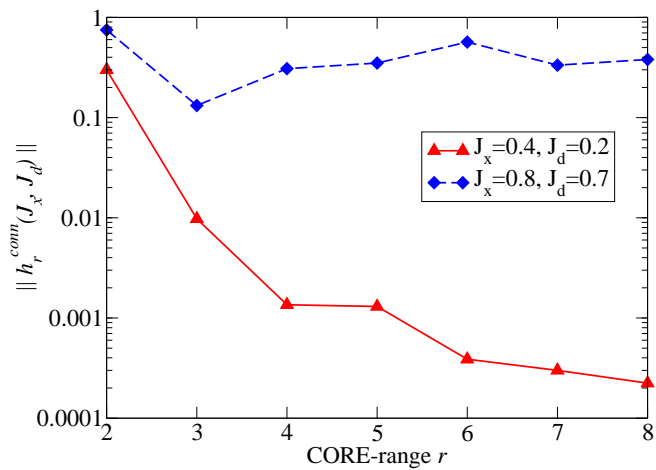}
\end{center}
\caption[]{(Color online) For $J_x=0.4$, $J_d=0.2$ the norm of $h_r^{conn}$ quickly decreases with increasing range $r$, whereas
for $J_x=0.8$, $J_d=0.7$ the norm of $h_r^{conn}$ has the same order of magnitude for ranges $2 \le r \le 8$.
\label{CORE_norm.fig} }
\end{figure}

On Fig.~\ref{CORE_norm.fig}, we plot the largest matrix element (in
absolute value) of the range-$r$ connected contribution
$h_{r}^{conn}$. For the coupling $J_x=0.4$, $J_d=0.2$, we indeed
observe a strong decrease of the amplitudes of the different processes
as a function of the range of interaction $r$. This gives us
confidence in the truncation beyond a certain range. However, the case
of $J_x=0.8$ and $J_d=0.7$ is an example where this CORE approach does not work, as will
be discussed below. Here the matrix elements of $h_{r}^{conn}$ remain
substantial even for large ranges $r$.

\begin{figure}
\begin{center}
\includegraphics[width=.48\textwidth,clip]{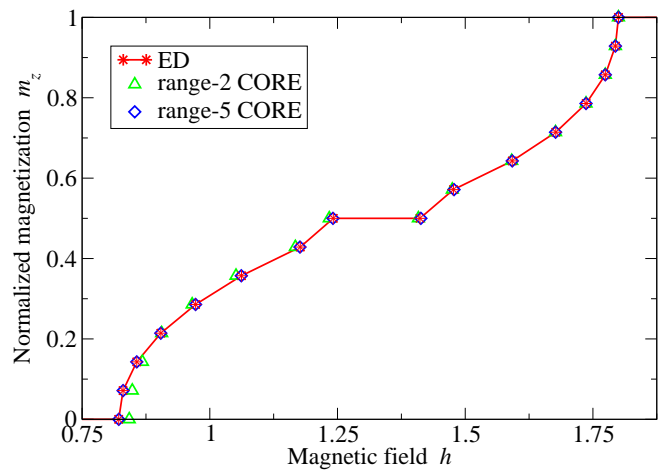}
\end{center}
\caption[]{(Color online) With increasing range, the CORE results converge towards the exact result. Parameters are: $2 \times L$ with $L=14$;
 $J_x=0.4$ and $J_d=0.2$. The effective model was obtained up to range 5, and then  solved on a 14-site chain.  
\label{CORE_convergence.fig} }
\end{figure}

For the cases where the matrix elements of $h_{r}^{conn}$ decrease fast with increasing range $r$,
we expect that CORE reproduces the low-energy physics of the system very well. In order to illustrate and strengthen this
point, we have \emph{exactly} solved different effective models obtained at a given truncation approximation.
On Fig.~\ref{CORE_convergence.fig}, we show for the example of $J_x=0.4$, $J_d=0.2$ that with increasing range,  
CORE results quickly converge to the exact ones.

Since  we have used the same basis for CORE as for perturbation theory in Sec. \ref{pert_theo.sec}, the range-2 effective Hamiltonian will 
also be a $t-V$ model  as in  Eq.~(\ref{ham_sf}). But for CORE, the dependence of $t$, $V$, and $\mu$ on the different interactions is different and reads:
\begin{eqnarray}
\label{t_V_CORE}
t = \frac {J_x - J_d} {2} & \qquad & V = E_{SS} + \frac {3} {2} J_{\perp} + \frac {J_x + J_d} {2} \\
\mu &=& -E_{SS} -h - \frac {J_{\perp}} {2} \nonumber
\end{eqnarray}
$E_{SS}$ is the ground state energy of the $2\times 2$ plaquette which is found to be:
\begin{eqnarray}
\label{E_SS}
E_{SS} &=& - \frac {J_{\perp}} {2} - \frac {J_x + J_d} {2} - \\
&-& \sqrt{ J_{\perp}^2 + J_x^2 + J_d^2 - J_x J_d - J_{\perp} (J_x + J_d) } \nonumber
\end{eqnarray}
Consequently, the condition $V/|t| > 2$ for the existence of a magnetization plateau at one half the saturation value translates into
\begin{equation}
\label{CORE_inequality}
\frac {1} {3-J_x} < J_d/J_x < \frac {3} {1+J_x}
\end{equation}
This criterion is shown on Fig.~\ref{phase_diag.fig} together with the
perturbative result.  For small interaction $J_x$, one observes that the plateau
phase boundaries given by Eqs.~(\ref{pert_theo_inequality}) and
(\ref{CORE_inequality}) practically coincide;  however, for increasing
interaction $J_x$, i.e. when $J_x$ can no more be treated as a small
perturbation, the two curves deviate from each other and CORE becomes more reliable.
Note that a range-2 CORE calculation is quite simple, can be done analytically by solving a $2\times 2$ plaquette and already improves the accuracy with respect
to perturbation theory. 

The limitations of this approach become evident for $J_x \sim J_d \ge
0.7$ where naive implementation of CORE fails to reproduce the absence
of magnetization plateaux and the matrix elements of $h_{r}^{conn}$ do not decrease
with increasing range $r$, as shown for $J_x=0.8$, $J_d=0.7$ in Fig.~\ref{CORE_norm.fig}. There is a simple symmetry argument to
understand why. For the specific case $J_x=J_d$, the Hamiltonian has
many additional symmetries, namely the exchange of the two spins on
\emph{any} given rung. Therefore, it can be shown that the effective
hoppings are strictly zero \emph{at all orders} in the CORE
approach, which of course leads to a gapped insulating phase at
half-filling. Indeed, in the CORE algorithm, it is
necessary that the ground-state has a finite overlap in the reduced
Hilbert space.

In the vicinity of the line $J_x=J_d$, this argument is
no longer strictly valid but, for practical calculations, we observe that in the CORE calculation, many overlaps become
very small, which results in effective models that are not accurate. Even if in principle, CORE could become accurate if
longer-range interactions are taken into account, we think that in such cases, CORE loses its practical utility.

 As we will discuss in the following, when the chosen
basis does not represent correctly the ground-state, practical
implementations of the CORE algorithm fail. When this is the case,
several strategies are possible~: (a) Keep other states for each
block; (b) Keep more states for each block or (c) Change the block. We
now turn to the discussion of each case by focusing on what are the
best block decomposition and/or states to keep.

\section{Choice of basis}
\label{choicebasis.sec}

CORE will be useful when one can keep a small number of states per
block and restrain to finite-range effective interactions. However,
checking the convergence is not always easy (except in 1D as we have
shown on Fig.~\ref{CORE_convergence.fig}) so that it would be useful
to have alternative information. In Ref.~[\onlinecite{Capponi2004}],
the authors proposed to used the reduced density matrix of the block
as a tool to compute the relative weights of each block state. This
procedure is similar to the density-matrix renormalization group method
(DMRG)~\cite{Schollwock2005} and can both help to choose the correct number of kept states,
or indicate that a given block decomposition might not be appropriate. 
Moreover, this analysis can be done independtly of CORE since it
relies on an \emph{exact} calculation and is rather easy since it can
be done on small clusters~\cite{noteA}.

\subsection{Rung density matrix}

\begin{figure}
\begin{center}
\includegraphics[width=.48\textwidth,clip]{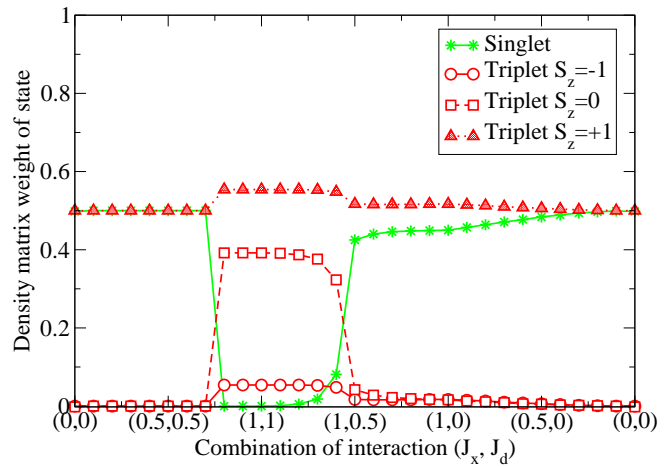}
\end{center}
\caption[]{(Color online) Reduced density matrix weights for the vertical dimer block obtained on a $2\times 12$ ladder as a 
function of $J_x$ and $J_d$ along a path in the phase diagram. 
Calculations have been carried out for a magnetization of $1/2$ of the saturation value.
\label{dens_vert_dimer.fig} }
\end{figure}

A first illustration of this reduced density matrix weights is given
on Fig.~\ref{dens_vert_dimer.fig}. We have considered a rung embedded
in a larger cluster and we trace out the other spins in the exact
density matrix obtained with the $m_z=\frac{1}{2} m_{sat}$ ground-state~\cite{noteB} (GS) $|\Psi\rangle$. By
diagonalizing this reduced density matrix, we obtain the probablity of
finding a certain block state, given that the overall system has a wavefunction
$|\Psi\rangle$. For our choice of block, we obtain 
the weights of the rung states, namely the singlet $|s\rangle$ and the 3 
triplets. We immediately observe that the singlet weight
vanishes or is very small when $J_x \sim J_d \ge 0.7$, which is
precisely the region where a naive CORE calculation fails to reproduce
the Haldane phase. In such cases, the exact GS has a vanishing or very small overlap in the naive CORE subspace, which explains the
failure of our previous approach. 

On Fig.~\ref{phase_diag.fig}, a line indicates the
region where the singlet weight becomes smaller than the $S_z=0$ triplet weight. It
corresponds precisely to the region with large $J_x\sim J_d$ where exact results
have shown that there is no magnetization plateau (see
Fig.~\ref{size_plat_ED.fig}). In this region, one needs to use other strategies for CORE. 

\subsection{Keeping 2 triplets}
In most of CORE approaches, the kept states have been taken as the
lowest in energy on a single block. This choice is of course natural
and is often the best one. However, in the case where both $J_x$ and
$J_d$ are close to $J_\perp$, we will argue that this is not the case. 
From the reduced density matrix weight
(Fig.~\ref{dens_vert_dimer.fig}), we have seen that the rung singlet
state does not describe accurately the \emph{exact} ground-state on a
large system. 
Therefore, a first 
modification to the standard CORE algorithm would be to keep the size of the truncated basis to 2 but take the largest-weight states, namely $|t_0\rangle$ and
$|t_{+1}\rangle$ in this region (the 3-fold degenerate triplets will have a Zeeman splitting in the presence of a magnetic field). 
As in the previous section, it is straightforward to compute \emph{analytically} the range-2
effective Hamiltonian that has again the form of $t$-$V$ model for spinless fermions as in Eq.~(\ref{ham_sf}) with 
\begin{eqnarray}
t&=&\frac{J_x+J_d}{2} \nonumber \\
V&=&\begin{cases}
E_{ss}+\frac{J_x+J_d-J_\perp}{2} & \mathrm{if} \hphantom{x} J_x \neq J_d \\
-t & \mathrm{if} \hphantom{x} J_x = J_d
\end{cases}
\end{eqnarray}
The crucial difference compared to the perturbative estimate of Eq.~(\ref{t_V_pert}) is that the effective hopping of
polarized triplets \emph{does not vanish} anymore when $J_x=J_d$. As a consequence, this effective mapping correctly 
predicts that $V/t<2$ so that there is no magnetization plateau. 

\begin{figure}
\begin{center}
\includegraphics[width=.48\textwidth,clip]{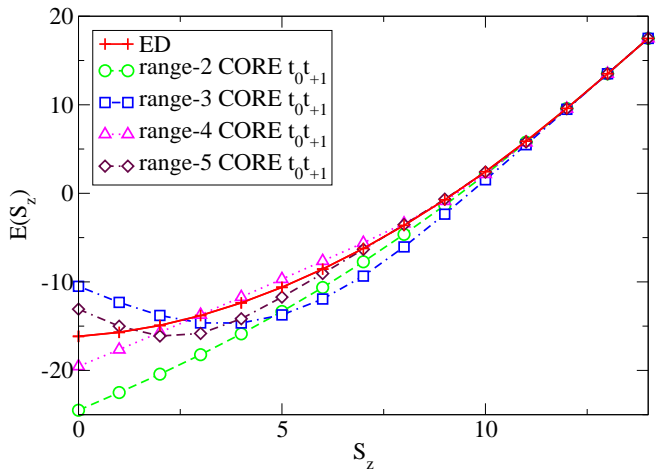}
\end{center}
\caption[]{(Color online) Comparison of energy vs magnetization for $J_x = J_d = 1$ on a $2\times14$ ladder given by ED 
and various CORE effective Hamiltonians obtained by choosing two triplets on each vertical rung and keeping up to range-$r$ interactions.
\label{e_vs_sz_tt_11.fig} }
\end{figure}

Of course, we can go beyond range-2 approximation by including larger
clusters in the CORE expansion. On Fig.~\ref{e_vs_sz_tt_11.fig}, we consider the highly non-perturbative case where all
couplings are equal and we 
observe a convergence of the CORE results towards ED data as we increase the range $r$. However, one needs to take into account longer-range
effective interactions in the small-magnetization region and even for range-5 CORE the agreement with ED data is not satisfactory.

\subsection{Keeping more states per block}
With the rung decomposition, we have observed that CORE is not very efficient if we restrict the truncated basis to singlet and polarized triplet.
We have obtained better results by keeping the two dominant triplet states but the convergence was still poor. 
The origin of these difficulty lies respectively in  
(a) the very low density matrix weight of some states (b) keeping states that are not the lowest in energy. 

\begin{figure}
\begin{center}
\includegraphics[width=.48\textwidth,clip]{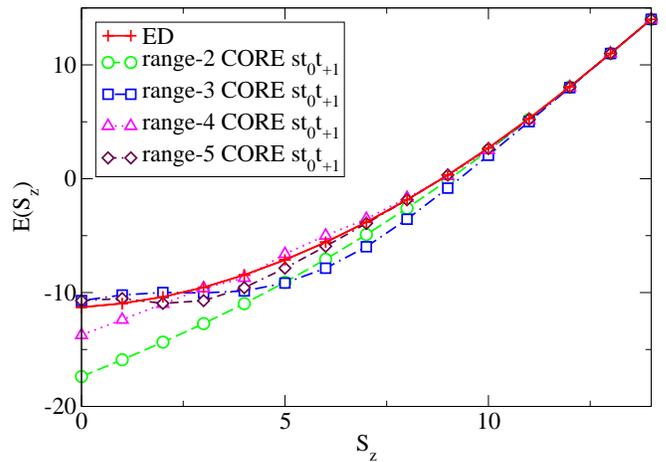}
\end{center}
\caption[]{(Color online) Comparison of energy vs magnetization for $J_x = 0.8$ and $J_d = 0.7$ on a $2\times14$ ladder given by ED 
and various CORE effective Hamiltonians obtained by choosing one singlet and two triplets on each vertical rung and keeping up to range-$r$ interactions.
\label{e_vs_sz_stt_87.fig} }
\end{figure}

To resolve these difficulties, we decide to keep all the rung states
except $|t_{-1}\rangle$. Clearly, if we increase the size of the CORE
subspace, we should get more accurate results but it will limit us in the possibility of studying such an effective model. 
On Fig.~\ref{e_vs_sz_stt_87.fig}, we have solved numerically various effective Hamiltonians obtained by keeping up to a given
range $r$ effective interactions. Although qualitatively correct since we do not observe any finite plateau in this region, the CORE
results converge slowly to the exact ones and give poor accuracy at small magnetization.

Note that in the case $J_x=J_d \geq 0.8$,  the rung singlet weight vanishes (as shown in Fig.~\ref{dens_vert_dimer.fig}) so that
the CORE results are identical whether we keep two triplets and the singlet, or only two triplets. 

As a conclusion for the rung basis, for a given block decomposition, we propose that an efficient CORE 
implementation should keep the first $M$ low-energy states per block 
 such that the total reduced density matrix weight is ``large''. 
Of course, in some cases, we had to keep a rather large part of the
total Hilbert space, which limits us both for numerical simulations
and for analytic study of the effective model. Therefore, another route can be
chosen by modifying the block decomposition.

\subsection{Horizontal dimer block}\label{hor_block.sec}

\begin{figure}
\begin{center}
\includegraphics[width=.48\textwidth,clip]{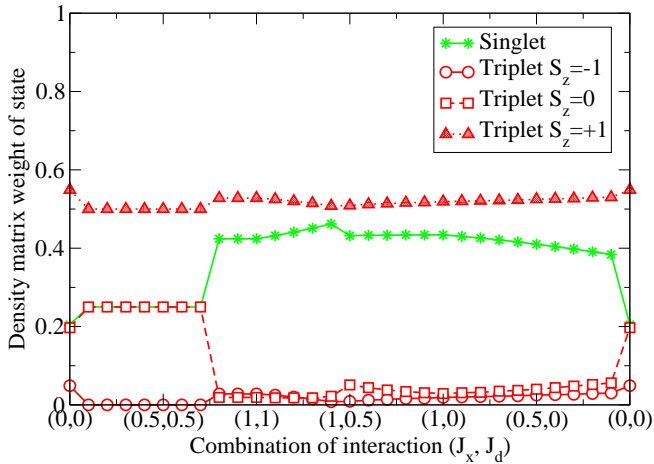}
\end{center}
\caption[]{(Color online) Reduced density matrix weights for the horizontal dimer block obtained on a $2\times 12$ ladder as a function of $J_x$
and $J_d$ along a path in the phase diagram. Calculations have been carried out for a magnetization of $1/2$ of the saturation value.
\label{dens_hor_dimer.fig} }
\end{figure}

Since the rung basis does not give accurate results in the strongly
frustrated regimes where all three couplings are of the same order,
we decide to choose another block decomposition with horizontal dimers,
keeping the singlet and polarized triplet: each block will again be
represented by a pseudo-spin 1/2. Fig.~\ref{dens_hor_dimer.fig} illustrates the reduced density matrix weights
 of the horizontal dimer states~\cite{noteB}.

By applying CORE, we obtain a new
effective ladder with many-body effective interactions. In its cluster
expansion, the CORE Hamiltonian should in principle contain all kind
of clusters interactions, including L-shape ones. However, because we
have to deal with \emph{connected} interactions, it can be shown that
some cancellations occur: for instance, since a given L-shape cluster
appears in only one rectangular-shape cluster, its contribution
exactly cancels out in the cluster expansion~\cite{noteC}. 
We have carried out a CORE calculation including up
to 6-rung interactions and we have solved these effective models
exactly.

\begin{figure}
\begin{center}
\includegraphics[width=.48\textwidth,clip]{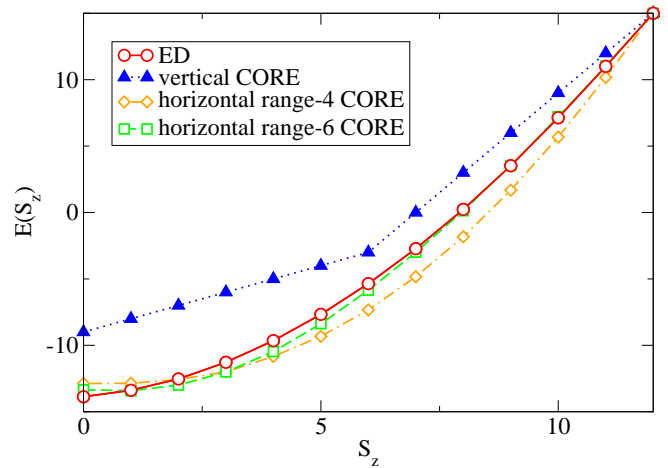}
\end{center}
\caption[]{(Color online) Energy vs magnetization obtained on a $2\times 12$ ladder with isotropic couplings $J_x=J_d=J_\perp=1$. 
CORE calculations are done with~: vertical rungs blocks, keeping $|s\rangle$ and $|t_{+1}\rangle$ and including \emph{all} effective interactions (corresponds to infinite range); horizontal dimer blocks, keeping $|s\rangle$ and $|t_{+1}\rangle$ and including range-4 and range-6 effective interactions. 
Exact results (ED) are shown for comparison.}
\label{e_vs_sz_vhrungs_111.fig}
\end{figure}

On Fig.~\ref{e_vs_sz_vhrungs_111.fig}, we compare range-4 and -6 calculations to exact results when $J_x=J_d=J_\perp=1$. Clearly, the horizontal dimer blocking scheme provides  very good agreement with exact data. In particular, we have a good convergence of CORE data when including longer-range effective interactions. Moreover, although the
rung basis wrongly predicted the existence of a magnetization plateau, here we observe a smooth energy vs magnetization curve and we have checked that there is no
indication of any plateau, as is known from exact calculations in this region.\\

At this stage, by choosing either a 2-site rung or leg blocking scheme, we are able to reproduce qualitatively the whole phase diagram for couplings $J_x$ and $J_d$
varying from 0 to $J_\perp$. This gives us confidence that CORE can be used to reduce the complexity of any microscopic model and still gives a correct description
of the properties in the presence of a magnetic field. For instance, the effective models that we have obtained can be solved exactly on clusters \emph{twice} as large
as for the original model. 

Because of the large flexibility in the choice of the block, we now turn to another decomposition of the ladder system. 

\subsection{Plaquette basis}
\begin{figure}
\begin{center}
\includegraphics[width=.48\textwidth,clip]{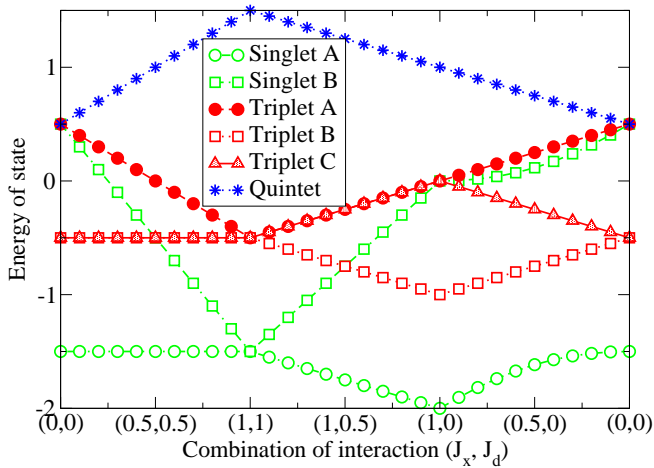}
\end{center}
\caption[]{(Color online) Energy of the plaquette states defined in Eq. (\ref{def_plaq_states}) as a function of the interaction along the ladder $J_x$ and the diagonal interaction $J_d$ (for $h=0$ and $J_{\perp}=1$).
\label{en_plaq_states.fig} }
\end{figure}

Another possible block decomposition of a ladder consists in 4-site plaquettes. We start the CORE algorithm by classifying its 16 states: 
two singlets, three triplets and one quintet. In the presence of a sufficiently strong magnetic field, only the polarized components will be
relevant so that we restrict ourselves to 6 states~: the fully polarized quintet
$| Q \rangle$, the three fully polarized triplets $| T_A \rangle$, $|
T_B \rangle$, and $| T_C \rangle$, and the two singlet states $| S_A
\rangle$, and $| S_B \rangle$. In order to make connection with the previous sections, we rewrite those states in terms of the
 rung basis ones:\\ 
the polarized quintet state $| S=2, S_z=2
\rangle$:
\begin{eqnarray}
| Q \rangle = | t_{+1} t_{+1} \rangle = | t_{+1} \rangle \otimes | t_{+1} \rangle {\rm;} \hphantom{lll} E_Q = \frac {1} {2} \big{(} J_{\perp} + J_x + J_d \big{)} \nonumber
\end{eqnarray}
the polarized triplet states $| S=1, S_z=1 \rangle$:
\begin{eqnarray}
| T_A \rangle = \frac {1} {\sqrt{2}} \big{(} | t_{0} t_{+1} \rangle - | t_{+1} t_{0} \rangle \big{)} {\rm;} && E_{T_A} = \frac {1} {2} \big{(} J_{\perp} - J_x - J_d \big{)} \nonumber \\
| T_B \rangle = \frac {1} {\sqrt{2}} \big{(} | s t_{+1} \rangle - | t_{+1} s \rangle \big{)} {\rm;} && E_{T_B} = \frac {1} {2} \big{(} J_{\perp} - J_x + J_d \big{)} \nonumber \\
| T_C \rangle = \frac {1} {\sqrt{2}} \big{(} | s t_{+1} \rangle + | t_{+1} s \rangle \big{)}  {\rm;} && E_{T_C} = \frac {1} {2} \big{(} J_{\perp} + J_x - J_d \big{)} \nonumber
\end{eqnarray}
the singlet states $| S=0, S_z=0 \rangle$:
\begin{eqnarray}
\label{def_plaq_states}
| S_A \rangle &=& | s s \rangle {\rm;} \hphantom{lll} E_{S_A} = - \frac {1} {2} \big{(} J_{\perp} + J_x + J_d \big{)}  - \gamma \nonumber \\
| S_B \rangle &=& \frac {1} {\sqrt{3}} \big{(} | t_{+1} t_{-1} \rangle + | t_{-1} t_{+1} \rangle - | t_{0} t_{0} \rangle \big{)} \hphantom{lllll} \nonumber \\
E_{S_B} &=& - \frac {1} {2} \big{(} J_{\perp} + J_x + J_d \big{)}  + \gamma \\
{\rm where} \hphantom{l} \gamma &=& \sqrt{ J_{\perp}^2 + J_x^2 + J_d^2 - J_{\perp} J_x - J_{\perp} J_d - J_x J_d } \nonumber
\end{eqnarray}
Note that, concerning the energies, the contribution of the magnetic field $h$ has been omitted. Fig.~\ref{en_plaq_states.fig} shows  the behavior 
of the energies as a function of $J_x$ and $J_d$ along  a path in the phase diagram. 

\begin{figure}
\begin{center}
\includegraphics[width=.48\textwidth,clip]{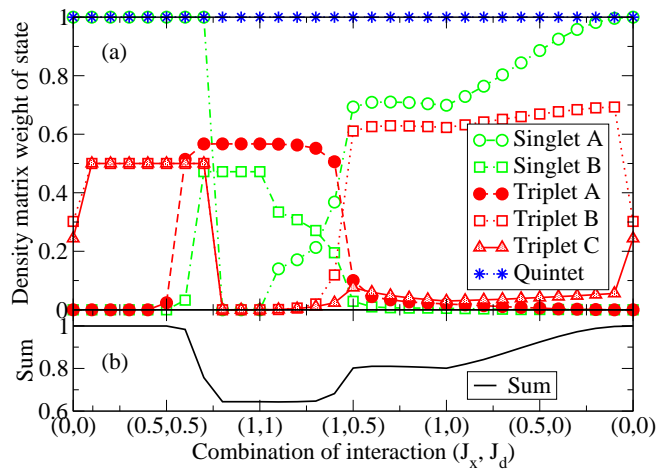}
\end{center}
\caption[]{(Color online) Reduced density matrix weights of the six chosen plaquette states (see Eq.~\ref{def_plaq_states}) as a function of  $J_x$ and $J_d$, 
obtained by ED from the exact ground-state of a $2\times 12$ ladder with various $S^{tot}_z$: 
(a) largest weight of the six plaquette states for all $S^{tot}_z$; (b) minimum over $S^{tot}_z$ of the cumulated weight of the six states.  
See text for details.
\label{dens_mat_states.fig} }
\end{figure}

We have previoulsy emphasized the use of the reduced density matrix weights in order to correctly choose which and how many block states should
be kept for a given CORE calculation. On Fig.~\ref{dens_mat_states.fig} (a), we plot the weights of our chosen plaquette states as a function 
of the ladder couplings (for each state, we compute its weight for all $S^{tot}_z$ and only plot the largest value); therefore, a small value indicates that a given plaquette
state is not relevant to describe the exact GS for \emph{any} $S^{tot}_z$, i.e. for \emph{any} magnetic field. Using this information, we can even reduce further the CORE basis
for some parameters~: if some weights are tiny or even zero (typically smaller 
than $5\%$),  we can reduce the CORE basis from 6 states to only 3 (one in each $S_z$-sector), and still have a good accuracy. Of course, reducing the CORE basis allows us to solve
exactly the effective model on much larger system sizes (up to $2\times 36$ for the original model).
 We can also use the reduced density matrix weights to check whether the 6-state basis correctly reproduces exact GS properties: on Fig.~\ref{dens_mat_states.fig} (b) is plotted the 
 total weight of these six states (we have computed the sum of these six weights for all $S^{tot}_z$ and we only show the smallest value). 
 Since this total weight is larger than $60\%$ in all the phase diagram (couplings and magnetic field), we expect that CORE effective
interactions should decay quickly with distance. 

\begin{figure}
\begin{center}
\includegraphics[width=.48\textwidth,clip]{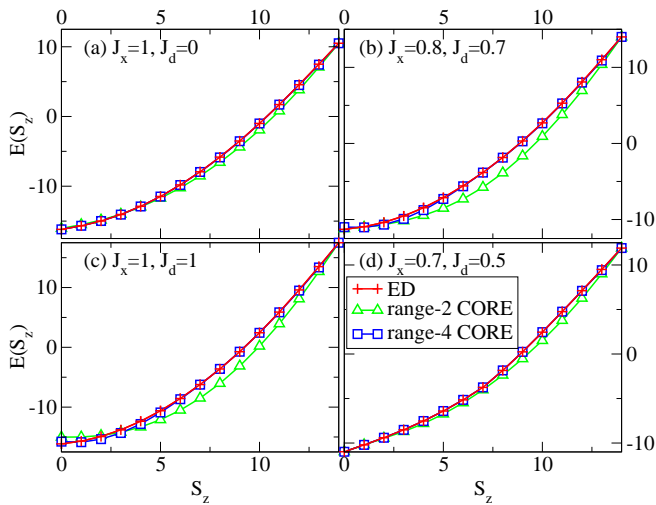}
\end{center}
\caption[]{(Color online) Comparison of energy vs magnetization given by ED and CORE Hamiltonians keeping up to range-2 or 4 effective interactions for a
$2\times 14$ ladder with various couplings. (a-c) for CORE, we keep only 3 states per plaquette, which have the largest reduced density matrix weights
(see Fig.~\ref{dens_mat_states.fig}); (d) for CORE, we keep six states per plaquette (see Eq.~(\ref{def_plaq_states})).
\label{e_vs_sz_pl_3st_10.fig} }
\end{figure}

\begin{figure}
\begin{center}
\includegraphics[width=.48\textwidth,clip]{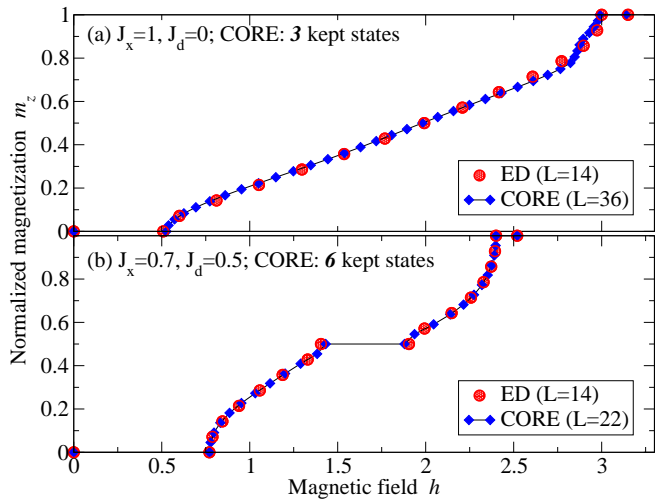}
\end{center}
\caption[]{(Color online) Comparison of magnetization curve for $2\times L$ ladder obtained with 
 ED and range-3 CORE calculations with (a) three kept plaquette states for $J_x = 1$ and $J_d = 0$ where there is no plateau; (b) 
six kept plaquette states for $J_x = 0.7$ and $J_d = 0.5$ where there is a finite plateau.
\label{mag_curve_pl_3st_10.fig} }
\end{figure}

Then, by solving exactly the effective models, we can compare the
energy vs $S_z$ curve to the exact one. On
Fig.~\ref{e_vs_sz_pl_3st_10.fig}, we plot our data obtained from
various couplings throughout the phase diagram; CORE calculations are
done either with 3 states (one per $S_z$ sector with the largest
reduced density matrix weight) or all six states according to
Fig.~\ref{dens_mat_states.fig}. In all cases, the CORE convergence is very
good and range-4 calculations are very accurate~\cite{noteD}. Being confident in
our effective models, we can compute the magnetization curve on much
larger systems and we show on Fig.~\ref{mag_curve_pl_3st_10.fig}
typical plots showing either the presence or absence of a magnetization
plateau at half-saturation, in full agreement with exact results. 
As a side remark, since our effective models are not necessarily particle-hole symmetric (in the bosonic language), 
we observe on Fig.~\ref{mag_curve_pl_3st_10.fig}(a) that the magnetization curve behaves differently close to $m_z\sim 0$ and $m_z\sim m_{sat}$; in particular, the precise
shape close to the saturation value converges quite slowly with the range $r$  of the effective CORE interactions.

As a conclusion for the frustrated ladder, we have been able to obtain
reliable effective Hamiltonians for various block decompositions
(vertical or horizontal dimers, or plaquette). For some parameters, some
choices are better than the others: for instance, in the
non-perturbative frustrated regime where all three couplings are of
the same magnitude, we have found that, with a plaquette decomposition
and keeping only three states per block (not always the low-energy ones), we had a quick convergence of
the CORE effective interactions that lead to a high accuracy, while CORE calculations are more difficult starting from rungs.  
The reduced density matrix criterion gives us two useful informations~: (i) it gives a systematic way to locate the energy cut-off and fix how many 
low-lying states per block should be kept in the CORE approach; (ii) it can also indicate that relevant block states are not necessarily the low-energy ones. 

\section{Two dimensional system}
\label{2d.sec}
\subsection{Anisotropic case}
\label{2d_anis.sec}
We now consider the 2-dimensional (2D) bilayer antiferromagnetic spin dimer XXZ model which is given by:
\begin{eqnarray}
\label{heisenberg_hamilton_2d}
  {\cal H}_{XXZ} &=&  - h \sum_{\alpha, i} S_{\alpha, i}^z + J \sum_{i} {\bf S}_{1,i} \cdot {\bf S}_{2,i} + \\
  &+& J' \sum_{\alpha, \langle i,j \rangle} \big{(} S_{\alpha, i}^x S_{\alpha, j}^x + S_{\alpha, i}^y S_{\alpha, j}^y +
\Delta S_{\alpha, i}^z S_{\alpha, j}^z \big{)}, \nonumber
\end{eqnarray}
where $\alpha = 1, 2$ denotes the layer index. The interlayer coupling $J$ is the largest one and has SU(2) symmetry, while the intra-layer coupling $J$ is taken with 
an anisotropy $\Delta$ ($\Delta=1$
corresponds to the isotropic $SU(2)$ case). Such a model has been recently introduced and studied numerically with a QMC technique~\cite{Ng2006a,Laflorencie2007}.
 As for the ladder model, it can be convenient to use the particle language, where the effective triplets behave as hardcore bosons that can have a solid 
(plateau region), superfluid (no plateau), or even a supersolid phase with both superfluid and solid order parameters. QMC simulations have shown that this model 
exhibits all these phases~\cite{Ng2006a}, including a large half-saturated magnetization plateau region. 

Analogously to the perturbation theory that has been applied on the
frustrated 2-leg ladder in section~\ref{pert_theo.sec}, we can carry
out a perturbation calculation, and then derive an effective 2D (hardcore) bosonic $t-V$ model~:
\begin{equation}\label{tV_boson.eq}
H_{tV} = t \sum_{\langle ij \rangle} \big{(} b_{i}^{\dagger} b_{j} + h. c.) + V \sum_{\langle ij \rangle} n_{i} n_{j}- \mu \sum_{i}n_{i}
\end{equation}
 or equivalently an effective XXZ spin-1/2 model. Again, we restrict ourselves to the singlet
$|s\rangle$ and the polarized triplet $|t_{+1}\rangle$ on each rung in order to describe the system in a magnetic field close to $m=1/2$.
 
\begin{figure}
\begin{center}
\includegraphics[width=.48\textwidth,clip]{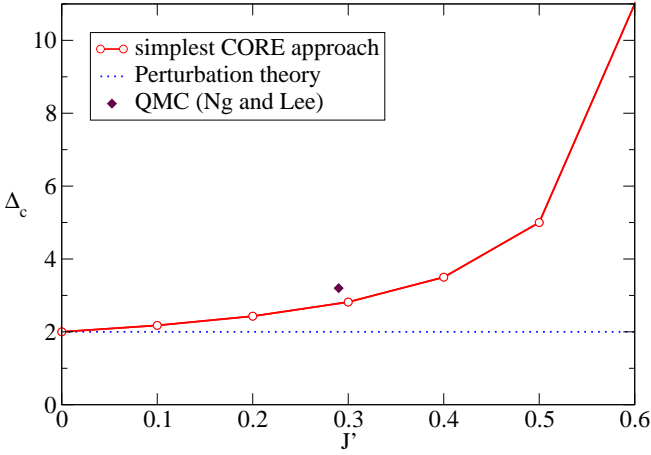}
\end{center}
\caption[]{(Color online) Critical anisotropy $\Delta_{c}$ above which there is a plateau phase as a function of $J'$ for perturbation theory and the simplest CORE approach for $J=1$. 
The QMC point: $J'/J=0.29$ and $\Delta_{c}=3.2$ is taken from Ref.~\onlinecite{Ng2006a}.
\label{anis_delta_crit_r2.fig} }
\end{figure}

One finds the following set of parameters:
\begin{equation}
\label{t_V_pert_2d}
t = \frac {J'} {2} \qquad V = \frac {\Delta J'} {2} \qquad \mu = J - h 
\end{equation}
This means that in perturbation theory,  one finds a finite plateau (superfluid-insulator
transition) when $V/|t| = \Delta > \Delta_c = 2$, which is independent of $J$ and $J'$ (for $J' \ne 0$).
This line is drawn in Fig.~\ref{anis_delta_crit_r2.fig} and  deviates from the QMC point.

Performing the simplest CORE approach we use the same two states
($|s\rangle$ and $|t_{+1}\rangle$).
Restricting to range-2 interactions, the 
parameters for the effective $t-V$ model can be easily derived:
\begin{eqnarray}
\label{t_V_CORE_2d}
t = \frac {J'} {2} & \qquad & V = \frac {\Delta J'} {2} + E_{GS} + \frac {3 J} {2} \\
\mu &=& - \frac {J} {2} - E_{GS} - h \nonumber
\end{eqnarray}
Here $E_{GS}$ is the ground state energy of the two dimers. We have also
calculated the reduced density matrix weight for this case~\cite{noteB} and
observed that the singlet and the polarized triplet represent more than 
$90\%$ of the total weight in the region of $0 \le J' \le 0.5$ and $0
\le \Delta \le 4$, which gives us confidence in the reliability of the effective model. 
We also noted that the weights depend only
marginally on $J'$ and $\Delta$ (data not shown). For instance, the weights for $\Delta=1$ can be retraced in
Fig.~\ref{dens_hb_frust.fig} in the sector where $J_d=0$.

 Comparing with the perturbative result of Eq.~(\ref{t_V_pert_2d}), one sees that $t$ remains unchanged whereas $V$ and $\mu$ are modified. Consequently, the criterion for a superfluid-insulator transition ($V/|t| = 2$) is no longer independent of $J$ and $J'$:
\begin{eqnarray}
\label{crit_CORE_2d}
\frac {V} {|t|} = 2 = \Delta_{c} + \frac {2 E_{GS} + 3 J} {J'}
\end{eqnarray}
This expression only coincides with the result of the perturbative
approach for $J'\rightarrow 0$. As is shown in Fig.~\ref{anis_delta_crit_r2.fig},  the critical value of the
anisotropy $\Delta_{c}$ increases monotonously with $J'$ if we set
$J=1$. Above this curve, the CORE approach predicts a solid phase which means
that the system exhibits a magnetization plateau at $1/2$ of its
saturation value. For comparison, we have added the QMC result~\cite{Ng2006a} indicating a solid phase (i.e. finite plateau) when 
$J'/J=0.29$ and $\Delta>3.2$, which 
is remarkably close to our simple estimate. 
Thus this CORE approach with two dimers which is
modest and very simple to carry out is a reliable tool to predict the occurence of  a
solid phase and gives a much better agreement with exact results than perturbation theory. 

Moreover, in the numerical study performed by Ng and
Lee~\cite{Ng2006a}, supersolid behaviour has been shown to occur close
to the solid phase. We believe that it would be desirable to derive
effective bosonic models showing such a rich phase diagram,
particularly for frustrated models which are not accessible by QMC
simulations due to the negative-sign problem. Our simple $t-V$ model
of Eq.~(\ref{ham_sf}) does not have a supersolid phase but exhibits
phase separation instead~\cite{Batrouni2000}. For that reason, we
extend the range of our CORE approach to range-4, i.e. we consider a
system of $2\times 2$ dimers~\cite{noteF}. 
We then obtain an effective hardcore bosonic model 
containing all possible interactions allowed by symmetry (i.e. conserving
the particle number)~:
\begin{eqnarray}
\label{heisenberg_hamilton_2d_t_V}
{\cal H} &=& \sum_{_i^l\square_j^k} \Big{[} \frac {C} {N} + \frac {\mu} {4} (\sum_i n_i) + \nonumber \\
&+& \frac {t_1^{(1)}} {2} (b_i^{\dagger} b_j + \circlearrowleft + h. c.) + t_1^{(2)} (b_i^{\dagger} b_k + b_l^{\dagger} b_j + h.c.) \nonumber \\
&+& t_2^{(1)} ( b_l^{\dagger} b_k^{\dagger} b_i b_j + h.c.) + t_2^{(2)} (b_l^{\dagger} b_j^{\dagger} b_i b_k + \circlearrowleft + h.c.) \nonumber \\
&+& t_1^{(3)} ( b_i^{\dagger} b_l (n_j + n_k - 2 n_j n_k) + \circlearrowleft + h.c.) \nonumber \\
&+& t_1^{(4)} ( b_i^{\dagger} b_k (n_j + n_l - 2 n_j n_l) + \circlearrowleft) \nonumber \\
&+& \frac {V_2^{(1)}} {2} (n_i n_j + \circlearrowleft) + V_2^{(2)} (n_i n_k + n_j n_l) \nonumber \\
&+& V_3^{(1)} ( n_i n_j (n_k + n_l -2 n_k n_l) + \circlearrowleft) \nonumber \\
&+& t_1^{(5)} (b_i^{\dagger} b_j  n_k n_l + \circlearrowleft +h.c.) + t_1^{(6)} (b_i^{\dagger} b_k  n_j n_l + \circlearrowleft)  \nonumber \\
&+& V_4 n_i n_j n_k n_l \Big{]}
\end{eqnarray}
Here we use the following notation: the sum $\sum_{_i^l\square_j^k}$ goes over the four plaquette sites. The circle $\circlearrowleft$ indicates that all contributions of one kind are included:
E.g. $\sum_{_i^l\square_j^k} \big{(} \frac {t_1^{(1)}} {2} ( b_i^{\dagger} b_j + \circlearrowleft )\big{)} = \frac {t_1^{(1)}} {2} ( b_i^{\dagger} b_j +
b_j^{\dagger} b_k + b_k^{\dagger} b_l + b_l^{\dagger} b_i )$.

\begin{figure}
\begin{center}
\includegraphics[width=.48\textwidth,clip]{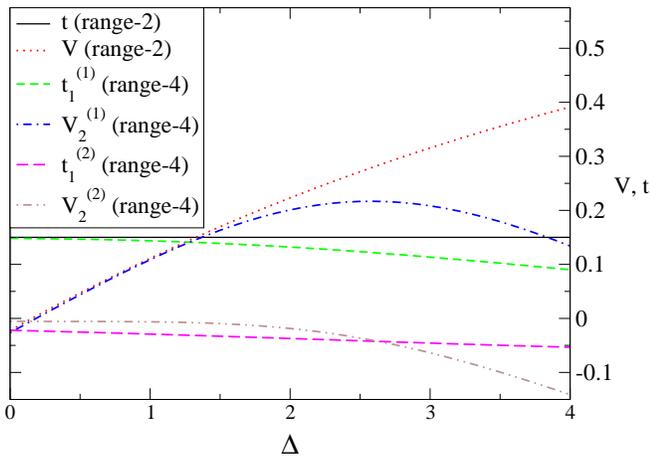}
\end{center}
\caption[]{(Color online) For fixed $J=1$ and $J'=0.3$: $t$ and $V$ as function of anisotropy $\Delta$ for range-2 CORE (see Eq.~(\ref{t_V_CORE_2d})); dominant parameters: 
$t_1^{(1)}$, $t_1^{(1)}$, $V_2^{(1)}$, and $V_2^{(2)}$ also as function of $\Delta$ for range-4 CORE (see Eq.~(\ref{heisenberg_hamilton_2d_t_V})).
\label{V_t_anis.fig} }
\end{figure}

In the region of interest, i.e. $J'/J=0.3$ and $0 \leq \Delta \leq 4$, although all parameters are non-vanishing, we find that 
the dominant terms are the following: the nearest and next-nearest neighbor hoppings
$t_1^{(1)}$ and $t_1^{(2)}=-t_1^{(6)}$, the nearest and next-nearest neighbor repulsion $V_2^{(1)}$ and $V_2^{(2)}$. 
The dependence of these parameters on the
anisotropy can be seen in Fig.~\ref{V_t_anis.fig} together
with the range-2 parameters $t$ and $V$ that have been
presented in Eq.~(\ref{t_V_CORE_2d}). For the range-2 data we note that the transition from the superfluid phase to the solid phase occurs
at $\Delta \sim 2.8$ for a inter-dimer coupling of $J'=0.3$.

Being able to compute a more refined effective Hamiltonian, one could
wonder about the possible effects of higher-order terms. This model is
very complicated and there is no simple criterion to predict the
existence or not of a plateau phase. Nevertheless, we believe that our range-4
effective Hamiltonian will also possess a plateau phase for large
$\Delta$. First of all, since this phase is gapped, we expect it to be robust to small additional interactions. 
Moreover, starting from a half-filled solid phase, the dominant longer-range terms have the following effect:
as shown in Fig.~\ref{V_t_anis.fig}, the diagonal density interaction $V_2^{(2)}<0$ enhances the stability of the solid phase, 
while the diagonal hoppings $t_1^{(2)}$ and $t_1^{(6)}$, that would favor a superfluid phase over the solid, remain small. 
This qualitative argument could be checked
quantitatively by QMC simulations, for instance with a simpler effective model
with no minus-sign problem. 

Moreover, the presence of diagonal single-particle hopping
$t_1^{(2)}$ and other terms should also stabilize a supersolid
phase~\cite{Sengupta2005}. In particular, \emph{removing a particle} from the half-filled solid creates a defect that can propagate
due to diagonal hopping, which results in a superfluid order parameter coexisting with the solid, {\it i.e.} a supersolid. 
On the contrary, if we \emph{add a particle} to the half-filled solid, our range-4 CORE effective Hamiltonian does not allow
for diagonal hopping since the two processes described by $t_1^{(2)}$ and $t_1^{(6)}$ in Eq.~(\ref{heisenberg_hamilton_2d_t_V}) 
 exactly cancel out ($t_1^{(2)}+ t_1^{(6)}=0$). From this observation, we expect that a supersolid phase is stable 
(respectively unstable)  
for filling smaller (respectively larger) than $1/2$, which is in perfect agreement with the findings of Ref.~\onlinecite{Ng2006a}.

It is also interesting to note that among
the longer-range effective interactions, correlated-hopping terms
could give rise to new phases~\cite{Bendjama2005} and could be
relevant for some geometries (e.g. Shastry-Sutherland lattice); here,
we have found that their amplitudes remain very small.

\subsection{Frustrated case}
\label{2d_frust.sec}
\begin{figure}
\begin{center}
\includegraphics[width=.48\textwidth,clip]{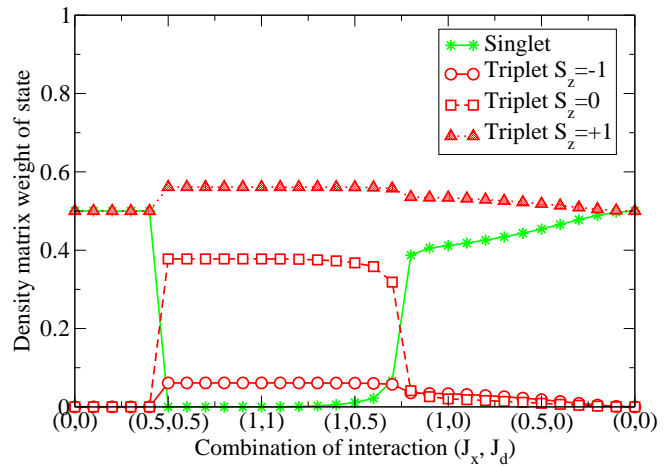}
\end{center}
\caption[]{(Color online) Reduced density matrix weights for the frustrated 2D Heisenberg bilayer obtained on a 
$2\times 16$ system as a function of $J_x$
and $J_d$ along a path in the phase diagram. Calculations have been carried out for a magnetization of $1/2$ of the saturation value.
\label{dens_hb_frust.fig} }
\end{figure}

Let us now regard the 2D antiferromagnetic bilayer model without anisotropy $\Delta$ but with a frustration $J_d$ between the two layers. 
This system corresponds to the 2D generalization of the frustrated Heisenberg ladder of Eq.~(\ref{heisenberg_hamilton}). Fig.~\ref{dens_hb_frust.fig} illustrates the reduced density matrix weights.\cite{noteB}

Carrying out a CORE calculation we can derive a $t-V$ model on two
dimers (see Eq.~(\ref{tV_boson.eq})) or a more complicated hardcore boson model including up to four dimers interactions of the form
of Eq.~(\ref{heisenberg_hamilton_2d_t_V}).

\begin{figure}
\begin{center}
\includegraphics[width=.48\textwidth,clip]{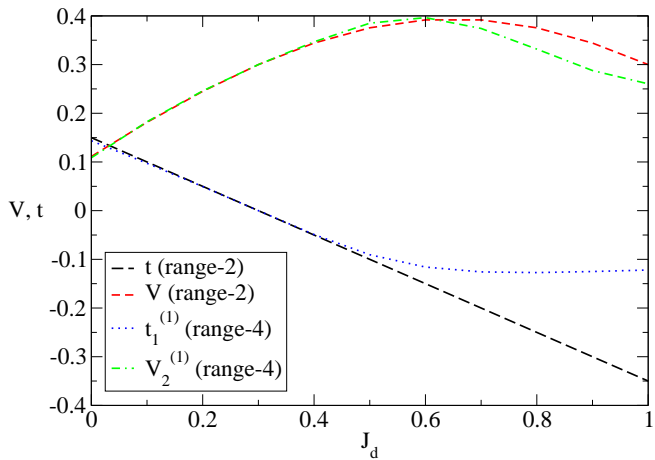}
\end{center}
\caption[]{(Color online) For fixed $J_x=0.3$ in the frustrated case: effective $t$ and $V$ parameters as a 
function of $J_d$ for range-2 CORE (see Eq.~(\ref{t_V_pert})); $t_1^{(1)}$ and $V_2^{(1)}$ also as function of $J_d$ for range-4 CORE (see Eq.~(\ref{heisenberg_hamilton_2d_t_V})).
\label{V_t_frust.fig} }
\end{figure}

In the range-4 CORE calculation, unlike in the previous anisotropic
case, we find that only the nearest-neighbor hopping $t_1^{(1)}$ and
repulsion $V_2^{(1)}$ are relevant since the other terms are at least
one order of magnitude smaller. These parameters are plotted in
Fig.~\ref{V_t_frust.fig}. Furthermore, one observes that for $J_x=0.3$
and $0 \leq J_d \leq 0.5$ the range-2 calculation has already
converged, i.e. the effective parameters are almost identical for
range-2 and range-4 approximations. Considering the large reduced density
matrix weights of the singlet and the
polarized triplet for these parameters (the sum of their contributions exceeds $95\%$ of the total weight, 
as shown in  Fig.~\ref{dens_hb_frust.fig}), 
 we expect that CORE
converges fast in this region.

Since the 2D or 1D superfluid-insulator transitions occur when $V/|t|=2$
which corresponds to a frustration of $J_d \sim 0.1$ for $J_x=0.3$, we
expect a plateau phase region similar to the one observed in the
ladder model (see Fig.~\ref{phase_diag.fig}).

Depending on $J_x$ and $J_d$ values, we find that higher-order terms may become important and could help stabilize a supersolid regime close to the plateau phase. A precise understanding
of the effects of these many-body effective interaction is, however, beyond the scope of our study. 

\section{Conclusion}
\label{conclusion.sec}

For frustrated ladders, naive perturbation in the rung basis 
completely fails to describe the disappearance of plateaux observed for large
 $J_x\sim J_d\sim J_\perp$. We have shown that, even with a simple range-2 CORE calculation keeping 2 states per rung, 
which can be done analytically, CORE gives much more accurate boundaries for the plateau region, as seen when comparing
exact data from Fig.~\ref{size_plat_ED.fig} and CORE predictions on Fig.~\ref{phase_diag.fig}. For the CORE calculations, we have found that it can be crucial
to use the information given by the exact reduced density matrix weights of the kept states in order to choose the best CORE 
basis, namely to answer the question: what is the best blocking scheme and how many block states should be kept ? 
Another advantage of the CORE calculation is that its accuracy can be
systematically improved by including longer-range effective interactions. By performing various block decompositions and 
keeping many-body effective interactions, we have shown that CORE is able to reproduce quantitatively the properties of
frustrated ladders in the presence of a magnetic field. The reduction of the Hilbert space allows us to solve exactly
the effective model on much larger system sizes  (up to $2\times 36$) compared to standard ED. 

We have also considered two-dimensional anisotropic or frustrated
Heisenberg bilayers. In the non-frustrated anisotropic case, our CORE
calculation improves over perturbation theory in locating the
condition for a plateau formation and is compatible with a recent QMC
study.  Fine details such as the occurence of supersolidity can also
be captured by computing longer-range effective interactions, such as
diagonal hopping. In particular, we find that assisted hopping is not
a relevant interaction for that geometry and these parameters.  The
frustrated bilayer was shown to have a similar effective model with
similar amplitudes. Therefore, we predict that it will have a similar
phase diagram as the anisotropic bilayer, containing superfluid, solid
and supersolid regions. The advantage of such a model is that it
respects $SU(2)$ symmetry and could be more realistic for materials
description. Physically, in the anisotropic case, the effective
repulsion between hardcore bosons increases with growing anisotropy,
leading to an insulating phase, whereas in the frustrated case 
the effective hopping is also strongly reduced with growing frustration.

\vspace*{1cm}
\noindent 
\acknowledgments
We thank IDRIS (Orsay, France) and CALMIP (Toulouse, France) for use of supercomputer
facilities.  We also thank the Agence Nationale de la Recherche (France) for support.  
We acknowledge fruitful discussions with F. Alet,  N. Laflorencie and A. L\"auchli. 

\bibliographystyle{aps}

\begin{thebibliography}{50}
\bibitem{noteA}Being a local object, we have checked that the local density
matrix is rather unsensitive to the total size of the system.
\bibitem{noteB}Since we focus on magnetization plateau at $m_{sat}/2$, we only show data obtained for
$m_z=\frac{1}{2} m_{sat}$, but we have checked that one finds similar results in its vicinity.
\bibitem{noteD}On a $2\times 14$ ladder, the largest Hilbert space dimension among all $S^{tot}_z$ is  
$40. 10^6$ for ED, but only 393 in our CORE subspace obtained by keeping 3 states per plaquette.
\bibitem{noteF}As we have explained in section~\ref{hor_block.sec}, L-shape clusters do not give any contribution in the CORE cluster expansion so that only
$p\times q$ rectangles should be considered for the effective interactions (see also Ref.~\onlinecite{noteC}).
\end{thebibliography}

\end{document}